# Implementation and Security Analysis of Practical Quantum Secure Direct Communication

Ruoyang Qi[1,*], Zhen Sun[2,*], Zaisheng Lin[2], Penghao Niu[1], Wentao Hao[2], Liyuan Song[4], Qin Huang[4], Jiancun Gao[1], Liuguo Yin[2,3,†], and Gui-Lu Long[1,3,5,6,†]

[1]State Key Laboratory of Low-Dimensional Quantum Physics and Department of Physics, Tsinghua University, Beijing 100084, China

[2]School of Information and Technology, Tsinghua University, Beijing, 100084, China

[3]Beijing National Research Center for Information Science and Technology, Beijing, 100084, China

[4]School of Electronic and Information Engineering, Beihang University, Beijing, 100191, China

[5]Innovative Center of Quantum Matter, Beijing,100084, China

[6]Beijing Academy of Quantum Information Science, Beijing, 100193, China

*These authors contributed equally to this work.

†Correspondence: LG Yin, E-mail: yinlg@tsinghua.edu.cn; GL Long, E-mail: gllong@tsinghua.edu.cn

ABSTRACT

Fast development of supercomputer and perspective quantum computer is posing increasing serious threats to communication security. Based on the laws of quantum mechanics, quantum communication offers provable security of communication, and is a promising solution to counter such threats. Quantum secure direct communication (QSDC) is one of the important branches of quantum communication. Different from other branches of quantum communication, it transmits secret information directly. Recently, remarkable progress has been made in the proof-of-principle experimental demonstrations of QSDC. However, it remains a technical feast to march QSDC into practical application. Here, we report an implementation of practical quantum secure communication system. The security is analyzed in the Wyner wiretap channel theory. The system uses a coding scheme based on concatenation of low density parity check (LDPC) codes, which works in a regime with realistic environment of high noise and high loss. The present system operates with a repetition rate of 1 MHz, and at a distance of 1.5 kilometers. The secure communication rate is 50 bps, which can effectively send text message and files such as image and sounds with a reasonable size.

# INTRODUCTION

Economic, political, and social well-being in the world depend crucially on secure communication infrastructures. Present communication is secured through encryption techniques, relying on pre-shared key and cryptographic protocols built from the computational difficulty of certain mathematical problems, for instance the RSA public key scheme[1]. There are potential dangers with the present secure communication system. On one hand, these cryptographic protocols are based on mathematical difficult problems that are not rigorously proven to have no efficient algorithms. These protocols may be broken one day, or might have been broken already privately, by some genius, as we do not yet know if efficient algorithm for solving these problems exists or not. On the other hand, with the fast development of super computers, and the perspective of practical quantum computer, some cryptography may become insecure[2]. Different from those cryptographic algorithms, physical-layer security is based on information theory rather than computational complexity. In 1975, Wyner presented the degraded wiretap channel model[3]. The secrecy capacity was defined as the supremum of all the achievable secure and reliable transmission rates. For classical communication, the estimation of the secrecy capacity in practical communication system is hard, because it is difficult for the legitimate parties to detect eavesdropping. When quantum systems, such as single photons or entangled photon pairs are used to transmit digital information, quantum physics principle gives rise to novel phenomena unachievable with classical transmission media[4]. It is impossible in principle to eavesdrop without disturbing the transmission to avoid detection. The pioneering work of Bennett and Brassard (BB84)[5] showed how to exploit quantum resources for random key agreement with high security using quantum key distribution technique. Quantum key distribution[5-9] distributes random key, rather than the information itself, and the information is sent through another classical communication.

In 2000, quantum secure direct communication (QSDC) was proposed, and it can transfer information directly without key distribution[10-12], which eliminates further security loopholes associated with key storage and ciphertext attacks[13,14], offering a new tool for selection in the zoo of secure communication protocols. Recently, experiments were reported of proof-of-principle demonstrations of QSDC based on single photons[15] and entanglement[16,17]. In particular, Zhang et al[17] demonstrated QSDC in a fiber with a distance of 500 meters using the two-step QSDC protocols[10,11].

Here, we report an experimental implementation of practical quantum secure communication system using a protocol based on the DL04 protocol[12]. To move QSDC forward into practical application, a number of key issues have to be solved. Security analysis of information transmission is crucial for practical application. According to Wyner's wiretap model, it is essential to let the system work with a capacity below the secrecy capacity of the channel. We estimated the secrecy capacity using the error rate from the sampling checking process of the system. Once this secrecy capacity estimation is completed, it is possible to design a coding scheme with a communication rate smaller than this secrecy capacity. We developed a coding scheme based on the concatenation of low density parity check (LDPC) codes[18,19], which is specifically designed for operating at the high loss and high error rate regime, unique for quantum communication. The platform shows that our implementation can effectively work in such realistic environment. In our system, the single photon source used is the attenuated faint laser pulse, and the repetition rate is 1 MHz. The distance is 1.5 kilometers, and the achieved secure information transmission rate is 50 *bps*, which is able to transmit text message and image or sound files of reasonable size.

**RESULTS**

**1）Practical DL04-QSDC (PDL-04 QSDC) protocol**

(1) Our practical quantum secure direct communication scheme is based on the DL04 QSDC protocol using single-photons[12]. The scheme is illustrated in details in Fig.1. It contains four steps. A legitimate information receiver, Bob, prepares a sequence of qubits, each randomly in one of the four states $|0\rangle$, $|1\rangle$, $|+\rangle$ and $|-\rangle$. Then he sends the sequence of states to the information sender, Alice. $|0\rangle$, $|1\rangle$ are the eigenstates of Pauli operator Z, and $|+\rangle$ and $|-\rangle$ are the eigenstates of Pauli operator X.

(2) Upon receiving the sequence of single photons, Alice chooses randomly some of them and measures them randomly in the Z-basis or the X-basis. She publishes the positions, the measuring-basis and measurement results of those single photons. Upon these information, Bob compares with his preparations of these states, and estimates the bit error rate of the Bob-to-Alice channel, and informs Alice through a broadcast channel. Thus, Alice can estimate the maximum secrecy capacity $C_s$ of the Bob-to-Alice channel using the wiretap channel theory. If the secrecy capacity is nonzero, then go to the next step. If the secrecy capacity $C_s$ is zero, they conclude the channel is insecure and terminate the process.

(3) Alice chooses a coding scheme with a transmission rate that does not exceed $C_s$ on the remaining qubits, to transmit secret information securely and reliably. She encodes a block of message to a code word with a LDPC code which will be described later. The basic operations are the following two unitary operations,

$$I = |0\rangle\langle 0| + |1\rangle\langle 1|, \quad Y = |1\rangle\langle 0| - |0\rangle\langle 1|,$$

mapping to '0' and '1' respectively, and they are further used for constructing the code words. Then she sends them back to Bob. To check eavesdropping in the channel from Alice to Bob, Alice also encodes some random numbers randomly in the encoded message sequence.

(4) Bob can deterministically decode Alice's message from his received signals by measuring the qubits in the same basis he prepared them. If the error rate is below the correcting capability of the LDPC code, the transmission is successful. Then they start from step (1) again to send another part of the secret message until they complete the transmission of the whole message. If the error rate is bigger than the correcting capability of the LDPC code, they stop the process.

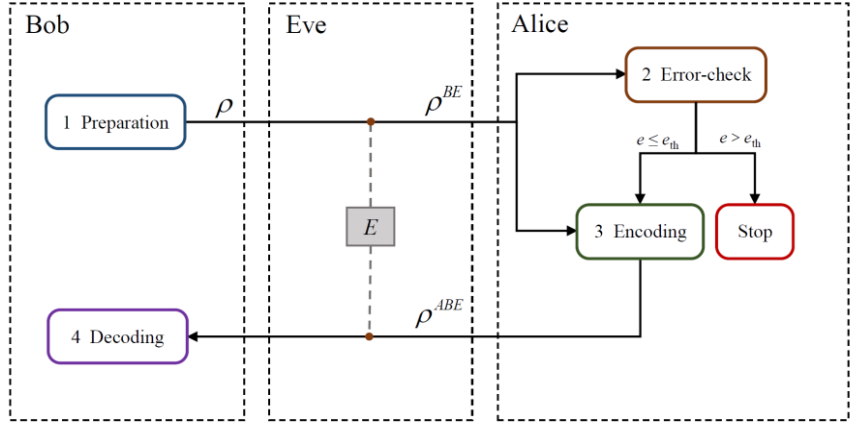

**Figure 1**. Illustration of the PDL04-QSDC protocol.

**2) Security analysis**

According to Wyner's wiretap channel theory[3], the secrecy capacity is

$$C_s = \max_{\{p\}}\{I(A:B) - I(A:E)\}, \tag{1}$$

where $I(A:B)$ and $I(A:E)$ are the mutual information between Alice and Bob, and Alice and Eve respectively. Moreover, $I(A:E)$ represents the maximum information that an eavesdropper can obtain, using the best strategy she can.

The state Bob prepared is a complete mixed state $\rho = (|0\rangle\langle 0| + |1\rangle\langle 1|)/2$, because he prepares with equal probability the four states, $|0\rangle$, $|1\rangle$, $|+\rangle$, $|-\rangle$. We consider the case of collective attack. The most general quantum operation which Eve may perform in the forward Bob-to-Alice channel consists of a joint operation on the qubit and some ancilla that belong to Eve,

$$\rho^{BE} = U(\rho \otimes |\varepsilon\rangle\langle\varepsilon|)U^+, \tag{2}$$

where $|\varepsilon\rangle$ represents Eve's ancillary state and $U$ is a unitary operation acting on the joint space of the ancilla and the qubit. Then, Eve resends the qubit to Alice and stores her ancilla until the qubit is sent back. Alice performs an operation, $I$ with probability $p$ or $Y$ with probability $1 - p$. After operating by Alice, the state becomes

$$\rho^{ABE} = p \cdot \rho_0^{BE} + (1-p) \cdot \rho_1^{BE}, \tag{3}$$

where $\rho_0^{BE} = I\rho^{BE}I$ and $\rho_1^{BE} = Y\rho^{BE}Y^+$. To gain Alice's information, Eve has to distinguish Alice's encoded qubit $\rho_0^{BE}$ from $\rho_1^{BE}$ by performing coherent measurements on any number of qubits and ancilla. The maximum mutual information between Alice and Eve is

$$I(A:E) \le \max_{\{U\}}\{S(\rho^{ABE}) - p \cdot S(\rho_0^{BE}) - (1-p) \cdot S(\rho_1^{BE})\}, \tag{4}$$

where $S(\rho)$ is the von Neumann entropy[20]. We obtain the maximum of mutual information between Alice and Eve (the detailed derivation is given in supplementary information),

$$I(A:E) \leq h(\xi), \tag{5}$$

where $\xi = \left(1 - \sqrt{(1-2p)^2 + (1-2e_x - 2e_z)^2[1-(1-2p)^2]}\right)/2$, $e_x$ and $e_z$ are the bit error rates in X-basis and Z-basis in the error-check respectively, and $h(x) = -x\log_2 x - (1-x)\log_2(1-x)$ is the binary Shannon entropy.

Because of imperfect efficiency of the detectors and channel loss, Bob cannot receive all the qubits. Gottesman has proven the security of Bennet-Brassard quantum key distribution protocol in the case where the source and detector are under the limited control of an adversary[21]. Similarly, considering the detectors and channel loss, the maximum mutual information between Alice and Eve becomes

$$I(A:E) \leq Q^{Eve} \cdot h(\xi), \tag{6}$$

where $Q^{Eve}$ is the maximum rate Eve can access the qubits.

The main channel can be modeled as a cascaded channel, which consists of a binary symmetric channel and a binary erasure channel in series[22], so that the mutual information between Alice and Bob is,

$$I(A:B) = Q^{Bob} \cdot [h(p + e - 2pe) - h(e)], \tag{7}$$

where $Q^{Bob}$ is the receipt rate in Bob's side, and $e$ is the bit error rate between Alice and Bob. We can estimate the lower bound of the secrecy capacity,

$$\begin{aligned} C_s &= \max_{\{p\}} \{I(A:B) - I(A:E)\} \\ &= \max_{\{p\}} \{Q^{Bob} \cdot [h(p+e-2pe) - h(e)] - Q^{Eve} \cdot h(\xi)\} \\ &= Q^{Bob} \cdot [1 - h(e)] - Q^{Eve} \cdot h(e_x + e_z) \\ &= Q^{Bob} \cdot [1 - h(e) - g \cdot h(e_x + e_z)] \end{aligned} \tag{8}$$

where $g$ represents the gap between $Q^{Eve}$ and $Q^{Bob}$, depending on the back channel loss and the efficiency of detector.

For any wiretap channel, if the secrecy capacity is non-zero, i.e., the legitimate receiver has a better channel than the eavesdropper, there exists some coding scheme that achieve perfect secrecy[23], such as two edge type LDPC codes based on coset encoding scheme[24].

**3) Experimental results**

We implemented the above scheme in fiber system with phase coding[25], the details of the experimental setup and methods are given in the material and method section. In our experiment, we initially set the distance at 1.5 km, which is a typical distance between different buildings of a secure area. Fig. 2 shows the error rates at Alice's and Bob's sites. In Alice's site, $e_x$ and $e_z$ are the error rates of measurements using the X-basis and Z-basis respectively. We estimate the error rate block by block, and each block contains 1312 × 830 = 1,088,960 pulses. In Fig.2, the horizontal axis is labeled against the number of blocks processed. Under normal working conditions, their values are about 0.8%. In Bob's site, of the pulses he sends to Alice previously, he receives 0.3% of them, namely for every 1000 pulses, 3 photon counts can be obtained when Bob measures the returned pulses. The error rate at Bob's site is smaller than those in Alice's site, due to the intrinsic robustness of the retrace-structure of light, usually about 0.6%. Here, the mean photon number is 0.1. The inherent loss of quantum channel is 14.5 dB, which includes the efficiency of the detector, about 70 %, and the optical elements, about 13 dB. Because the mean photon number is 0.1 and the channel loss of 1.5 kilometers fiber is 0.6 dB, this gives the total loss of the system as 25.1 dB. Shown in Fig.3, the mutual information $I(A:B)$ and $I(A:E)$ versus the loss of the system are two straight lines. The area between these two lines are the information-theoretic secure area, namely, if a coding scheme with an information rate within these areas, it will guarantee the security reliably. The error rates are set at values as above, namely $e$ is 0.6 %, $e_x$ and $e_z$ are 0.8 %. The secrecy capacity can be estimated as 0.00184 for loss at 25.1 dB. For the number of $N$ in the pseudo-random sequence, we set it to $N = 830$, after optimization. Together with the chosen LDPC code, our coding scheme gives a transmission rate 0.00096, when the bit error rate is chosen as $10^{-6}$. Meanwhile, $I(A:E) = g \cdot Q^{Bob} \cdot h(e_x + e_z) = 9.1 \times 10^{-4}$, where the loss of the back channel, includes the efficiency of the detector and channel loss, is about 4.1 dB, so that $g = 2.57$. This gives a secure information rate of 50 $bps$, which is well-within the secure area in Fig.3.

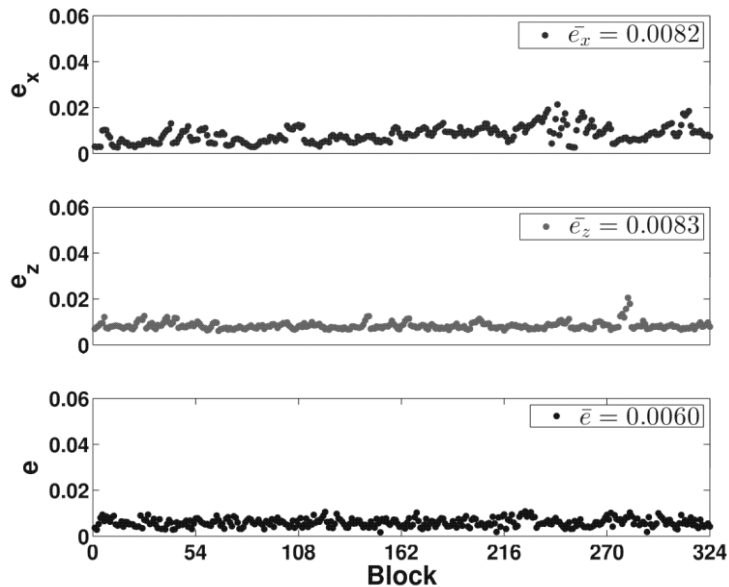

**Figure 2**. System stability with different message block. $e_x$ and $e_z$ are the error rates of measurements using the X-basis and Z-basis respectively, in Alice's site. $e$ is the error rate at Bob's site. We estimate the error rate block by block, and each block contains 1312×830 pulses. The mean photon number is 0.1. The inherent loss of quantum channel is 14.5 dB, which includes the efficiency of the detector, about 70 %, and the optical elements, about 13 dB. The total loss of the system is 25.1 dB, with distance of 1.5 kilometers.

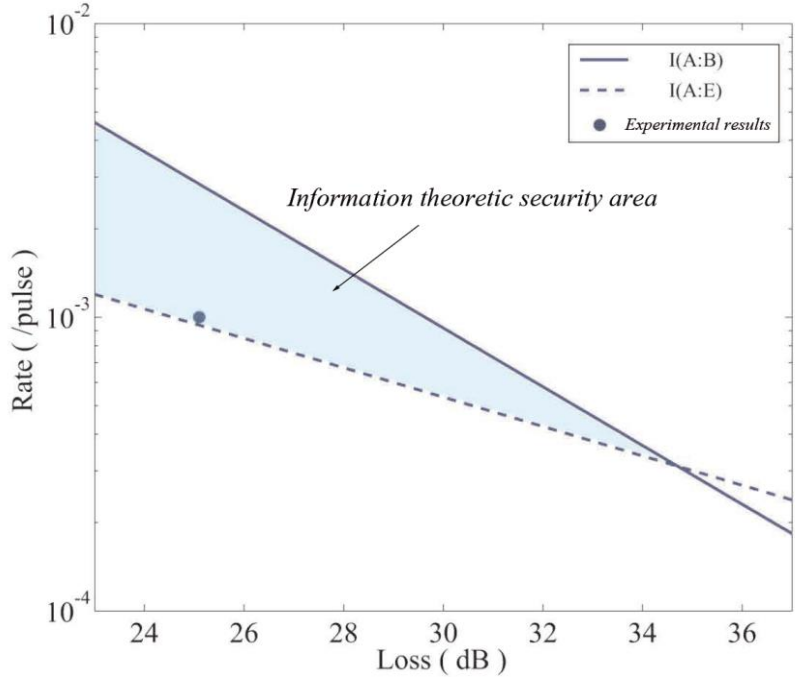

**Figure 3**. The solid line represents the mutual information between Alice and Bob, the capacity of main channel. For secure transmission, the information rate cannot exceed it. The dotted line is the mutual information between Alice and Eve, the maximum information that an eavesdropper can obtain. The error rates are set at values as above, namely $e$ is 0.6 %, $e_x$ and $e_z$ are 0.8 %. Symbols represent experimental results. We set the length of pseudo-random sequence as 830. Together with the chosen LDPC code, our coding scheme gives a transmission rate 0.00096 when the bit error rate is under $10^{-6}$. As it is greater than the mutual information between Alice and Eve, both the security and reliability of the information transmission are assured.

## DISCUSSION

It is well-known that in quantum communication, the photon loss is tremendous, due to inefficient photon source, high channel loss and low efficiency detector. Traditional error-correcting coding is usually designed for working in the low-loss and low-noise regime, and it cannot be used for our purpose. To guarantee the reliability and security of transmission for QSDC, we designed a coding scheme based on the concatenation of Low-Density Parity-Check (LDPC) codes, and pseudo random sequence, with the preprocessing based on the universal hashing families (UHF)[26]. For each message block ***m***, Alice generates a sequence of random bits, which is denoted as ***r***, of length $k$. Then, she maps (***m***, ***r***) to a new vector ***u*** by UHF, and encodes the vector ***u*** to ***v*** of length $l$, using the generator matrix of a specified LDPC code. Alice maps each coded bit to sequence of length $N$ to obtain a transmitted sequence of length $Nl$, which is transmitted over the quantum channel. After receiving the modulated pulses from Alice, Bob makes measurement in the same basis as he prepared them. Though only a fraction of photons in a pseudo random sequence can reach Bob's site, he can still read out the coded bit by looking at the log-likelihood ratios of each coded bit calculated from the received sequence. Then, he decodes the LDPC code by an iterative propagation decoding algorithm with the log-likelihood ratios. Information theoretic security can be guaranteed[27], as long as $I(A:E) \leq k/(N \cdot l)$.

## MATERIALS AND METHODS

The experimental setup is shown in Fig. 4. Bob prepares a sequence of single photon pulses. After polarization control and attenuation, it goes to the Mach-Zehnder ring where a random phase of the value of 0, π/2, π and 3π/2, is encoded, which is equivalent to preparing qubit randomly in $|0\rangle$, $(|0\rangle+|1\rangle)/\sqrt{2}$, $|1\rangle$ and $(|0\rangle-|1\rangle)/\sqrt{2}$ state respectively. Then it is sent to Alice's site through a fiber of 1.5 km long. After arriving at Alice's site, it is separated into two parts, one goes to the encoding module, and the other goes to the control module. In the control module, the qubits are measured, and the results are compared with Bob through the classical communication line connecting the two FPGAs shown at the bottom of Fig.4. At the same time, encoding is performed in the encoding-module. If the error rate is smaller than the threshold, the encoding part is allowed to send the single photons back to Bob, through the same fiber, and then are guided to the single photon detectors where they are measured. The three phase modulators, the single photon detectors, and the encoding of messages are controlled by the FPGAs at the two sites, which are further controlled by upper-position computers.

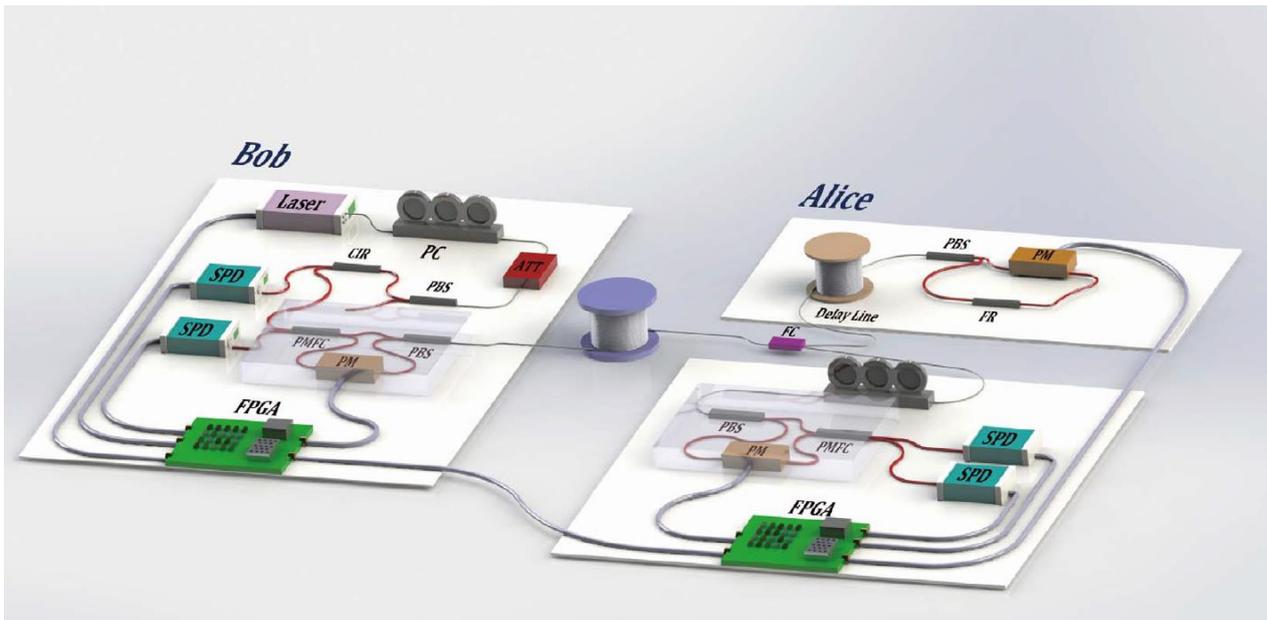

**Figure 4**. Experiment setup. A strongly attenuated 1550 nm laser is used as an approximate single-photon source with systematic pulse repetition frequency of 1 MHz. Bob sends the single photons in a superposition of two time-bins with a relative phase to Alice, and Alice randomly chooses one of two possible tasks, error-check or coding. Both sides are each controlled by a field programmable gate array (FPGA), and the operation of the four states of single photon is realized with a commercial lithium niobate modulator (PM, phase modulator). PC, polarization controller. PBS, polarization beam splitter. ATT, attenuator. CIR, optical circulator. FC, fiber coupler. SPD, Single photon detector. PMFC, polarization maintaining filter coupler. FR, Faraday Rotator.

The advantage of such forward-backward routing of the photon pulses is the automatic compensation of drift of polarizations of the time-bin pulses, because they exchange their routes after reflected by the Faraday rotator at Alice's site. This automatic compensation design was proposed by Martilelli[28], and has also been used in the plug-play QKD system[29]. The major difference between the plug-play QKD[29] scheme and DL04-based schemes, such as Refs. 7,12,15 and this PDL04-QSDC scheme is in the strength of light pulses in the forward channel. In Refs. 7,12,15, single photons are used in both the forward and backward channels, whereas in plug-play QKD[29] the forward channel uses strong classical light pulse, only the Alice to Bob backward channel uses single photon pulses. This mechanism of automatic compensation of polarization fluctuation enhances greatly the interference and leads to high visibility[30]. However, in the check-module of our system, such retrace-light circuit is not applicable, and active polarization, namely one monitors the drift and forcibly restore them when it reaches a value, has to be used. As a result, the error rate in the check-mode is usually higher than that in the communication-mode.

In summary, we have implemented a full practical quantum secure direct communication system with realistic environment of high noise and high loss. To combat error and loss, LDPC code and pseudo-random sequence techniques are applied. The security of the system is analyzed in details using the wiretap channel theory. Given the error rates, the secrecy capacity of the channel can be estimated. When the secrecy capacity is nonzero, a coding scheme with an information rate less than the secrecy capacity will ensure both the security of the information transmission and reliability of the information transmission. At a practical meaningful distance of 1.5 kilometers, a secure information rate of 50 *bps* is achieved. These parameters are premature, and there is much room for improvement. With current technology, an information rate of a dozens of *kbps* is achievable.

## ACKNOWLEDGMENTS

This work was supported by the National Basic Research Program of China under Grant Nos. 2017YFA0303700 and 2015CB921001, National Natural Science Foundation of China under Grant Nos. 61727801, 11474181 and 11774197. This work is supported in part by the Beijing Advanced Innovation Center for Future Chip (ICFC).

## CONFLICT OF INTERESTS

The authors declare that they have no conflict of interest.

## CONTRIBUTONS

RYQ, ZSL, PHN, JCG and GLL designed the protocol and the optical circuits, and setup the physical layout. ZS,WTH, LYS, QH, LGY made the LDPC coding and pseudo-M series. RYQ, ZSL, LGY and GLL completed the security analysis. LGY and GLL supervised the project. GLL leads the whole project. All authors contributed to the writing of the paper.

## REFERENCES

1   Rivest, R. L., Shamir, A. & Adleman, L. A method for obtaining digital signatures and public-key cryptosystems. *Communications of the ACM* **21**, 120–126 (1978).

2   Shor, P. W. Algorithms for quantum computation: Discrete logarithms and factoring. In *Foundations of Computer Science, 1994 Proceedings., 35th Annual Symposium on*, 124–134 (Ieee, 1994).

3   Wyner, A. D. The wire-tap channel. *Bell Labs Technical Journal* **54**, 1355–1387 (1975).

4   Gisin, N., Ribordy, G., Tittel, W. & Zbinden, H. Quantum cryptography. *Reviews of Modern Physics* **74**, 145 (2002).

5   Bennet, C. H. Quantum cryptography: Public key distribution and coin tossing. In *Proc. of IEEE Int. Conf. on Comp., Syst. and Signal Proc., Bangalore, India, Dec. 10-12, 1984* (1984).

6   Ekert, A. K. Quantum cryptography based on bell's theorem. *Physical Review Letters* **67**, 661(1991).

7   Deng, F. G., Long, G. L. Bidirectional quantum key distribution protocol with practical faint laser pulses. *Physical Review A*, **70**, 012311 (2004).

8   Lucamarini, M. & Mancini, S. Secure deterministic communication without entanglement. *Physical Review Letters* **94**, 140501 (2005).

9   Beaudry, N. J., Lucamarini, M., Mancini, S. & Renner, R. Security of two-way quantum key distribution. *Physical Review A* **88**, 062302 (2013).

10  Long, G.-L. & Liu, X.-S. Theoretically efficient high-capacity quantum-key distribution scheme. *Physical Review A* **65**, 032302 (2002).

11  Deng, F.-G., Long, G. L. & Liu, X.-S. Two-step quantum direct communication protocol using the einstein-podolsky-rosen pair block. *Physical Review A* **68**, 042317 (2003).

12  Deng, F.-G. & Long, G. L. Secure direct communication with a quantum one-time pad. *Physical Review A* **69**, 052319 (2004).

13  Niu, P.-H. *et al*. Measurement-device-independent quantum communication without encryption. *Science Bulletin* (2018), doi.org/10.1016/j.scib.2018.09.009

14  Zhou, Z.-R., Sheng, Y.-B., Niu, P.-H., Yin, L.-G. & Long, G.-L. Measurement-device-independent quantum secure direct communication. *arXiv preprint arXiv:1805.07228* (2018)

15  Hu, J.-Y. *et al*. Experimental quantum secure direct communication with single photons. *Light: Science & Applications* **5**, e16144 (2016).

16  Zhang, W. *et al*. Quantum secure direct communication with quantum memory. *Physical Review Letters*

**118**, 220501 (2017).

17 Zhu, F., Zhang, W., Sheng, Y. & Huang, Y. Experimental long-distance quantum secure direct communication. *Science Bulletin* **62**, 1519–1524 (2017).

18 Chen, Z., Yin, L., Pei, Y. & Lu, J. Codehop: physical layer error correction and encryption with ldpc-based code hopping. *Science China Information Sciences* **59**, 102309 (2016).

19 Wang, P., Yin, L. & Lu, J. Efficient helicopter- satellite communication scheme based on check-hybrid ldpc coding. *Tsinghua Science and Technology* **23**, 323–332 (2018)

20 Holevo, A.S. Bounds for the quantity of information transmitted by a quantum communication channel. *Problemy Peredachi Informatsii* **9**, 3–11 (1973).

21 Gottesman, D., Lo, H., Lutkenhaus, N. & Preskill, J. Security of quantum key distribution with imperfect devices. *Quantum Information & Computation* **4**, 325–360 (2004).

22 MacKay, D. J. & Mac Kay, D. J. *Information theory, inference and learning algorithms* (Cambridge university press, 2003).

23 Thangaraj, A., Dihidar, S., Calderbank, A. R., McLaughlin, S. W. & Merolla, J.-M. Applications of ldpc codes to the wiretap channel. *IEEE Transactions on Information Theory* **53**, 2933–2945 (2007).

24 Rathi, V., Andersson, M., Thobaben, R., Kliewer, J. & Skoglund, M. Performance analysis and design of two edge-type ldpc codes for the bec wiretap channel. *IEEE Transactions on Information Theory* **59**, 1048–1064 (2013).

25 Brendel, J., Gisin, N., Tittel, W. & Zbinden, H. Pulsed energy-time entangled twin-photon source for quantum communication. *Physical Review Letters* **82**, 2594 (1999).

26 Carter, J. L. & Wegman, M. N. Universal classes of hash functions. *Journal of Computer and System Sciences* **18**, 143–154 (1979).

27 Tyagi, H. & Vardy, A. Universal hashing for information-theoretic security. *Proceedings of the IEEE* **103**, 1781–1795 (2015).

28 Martinelli, M. A universal compensator for polarization changes induced by birefringence on a retracing beam. *Optics Communications* **72**, 341-344 (1989).

29 Muller, A. *et al*. "plug and play" systems for quantum cryptography. *Applied Physics Letters* **70**, 793–795 (1997).

30 Sun, S.-H., Ma, H.-Q., Han, J.-J., Liang, L.-M. & Li, C.-Z. Quantum key distribution based on phase encoding in long-distance communication fiber. *Optics Letters* **35**, 1203–1205 (2010).

# Supplementary Information
## Implementation and Security Analysis of Practical Quantum Secure Direct Communication

The maximum mutual information between Alice and Eve can be calculated as follows

$$I(A:E) \leq \max_{\{U\}}\{S(\rho^{ABE}) - p \cdot S(\rho_0^{BE}) - (1-p) \cdot S(\rho_1^{BE})\},$$

where

$$\rho^{ABE} = p \cdot \rho_0^{BE} + (1-p) \cdot \rho_1^{BE}$$
$$= p \cdot U(\rho \otimes |\varepsilon\rangle\langle\varepsilon|)U^+ + (1-p) \cdot YU(\rho \otimes |\varepsilon\rangle\langle\varepsilon|)U^+Y^+$$

Since $\rho_0^{BE}$ and $\rho_1^{BE}$ only differs from $\rho \otimes |\varepsilon\rangle\langle\varepsilon|$ by some unitary transformation,

$$S(\rho_0^{BE}) = S(\rho_1^{BE}) = S(\rho) = 1.$$

Hence,

$$I(A:E) \leq \max_{\{U\}}\{S(\rho^{ABE})\} - 1.$$

The effect of the unitary operation may be represented as

$$U|0\rangle|\varepsilon\rangle = |0\rangle|\varepsilon_{00}\rangle + |1\rangle|\varepsilon_{01}\rangle = |\varphi_1\rangle$$
$$U|1\rangle|\varepsilon\rangle = |0\rangle|\varepsilon_{10}\rangle + |1\rangle|\varepsilon_{11}\rangle = |\varphi_2\rangle$$
$$YU|0\rangle|\varepsilon\rangle = |1\rangle|\varepsilon_{00}\rangle - |0\rangle|\varepsilon_{01}\rangle = |\varphi_3\rangle$$
$$YU|1\rangle|\varepsilon\rangle = |1\rangle|\varepsilon_{10}\rangle - |0\rangle|\varepsilon_{11}\rangle = |\varphi_4\rangle$$

Unitarity is guaranteed if the following conditions are satisfied,

$$\langle\varepsilon_{00}|\varepsilon_{00}\rangle + \langle\varepsilon_{01}|\varepsilon_{01}\rangle = 1$$
$$\langle\varepsilon_{11}|\varepsilon_{11}\rangle + \langle\varepsilon_{10}|\varepsilon_{10}\rangle = 1 .$$
$$\langle\varepsilon_{00}|\varepsilon_{10}\rangle + \langle\varepsilon_{01}|\varepsilon_{11}\rangle = 0$$

The corresponding Gram matrix of $\rho^{ABE}$ is explicitly written as [1],

$$G = \frac{1}{2}\begin{pmatrix} p & 0 & 2i\alpha\sqrt{p(1-p)} & \sqrt{p(1-p)}\delta \\ 0 & p & -\sqrt{p(1-p)}\delta^* & -2i\beta\sqrt{p(1-p)} \\ -2i\alpha\sqrt{p(1-p)} & -\sqrt{p(1-p)}\delta & 1-p & 0 \\ \sqrt{p(1-p)}\delta^* & 2i\beta\sqrt{p(1-p)} & 0 & 1-p \end{pmatrix}$$

where $\alpha = \text{Im}(\langle\varepsilon_{01}|\varepsilon_{00}\rangle)$, $\beta = \text{Im}(\langle\varepsilon_{10}|\varepsilon_{11}\rangle)$, $\delta = \langle\varepsilon_{01}|\varepsilon_{10}\rangle - \langle\varepsilon_{00}|\varepsilon_{11}\rangle$.

The eigenvalues of $G$ are given by

$$\lambda = \frac{1}{4} \pm \frac{1}{2}\sqrt{p(1-p)(\Delta_1 \pm \Delta_2)^2 + (p - \frac{1}{2})^2}$$

where $\Delta_1 = \sqrt{(\alpha - \beta)^2}$, $\Delta_2 = \sqrt{(\alpha + \beta)^2 + \delta\delta^*}$.

$$U|+\rangle|\varepsilon\rangle = \frac{1}{\sqrt{2}}(|0\rangle|\varepsilon_{00}\rangle+|1\rangle|\varepsilon_{01}\rangle) + \frac{1}{\sqrt{2}}(|0\rangle|\varepsilon_{10}\rangle+|1\rangle|\varepsilon_{11}\rangle)$$

$$= |+\rangle\frac{|\varepsilon_{00}\rangle+|\varepsilon_{01}\rangle+|\varepsilon_{10}\rangle+|\varepsilon_{11}\rangle}{2} + |-\rangle\frac{|\varepsilon_{00}\rangle-|\varepsilon_{01}\rangle+|\varepsilon_{10}\rangle-|\varepsilon_{11}\rangle}{2}$$

$$= |+\rangle|\varepsilon_{++}\rangle + |-\rangle|\varepsilon_{+-}\rangle$$

$$U|-\rangle|\varepsilon\rangle = \frac{1}{\sqrt{2}}(|0\rangle|\varepsilon_{00}\rangle+|1\rangle|\varepsilon_{01}\rangle) - \frac{1}{\sqrt{2}}(|0\rangle|\varepsilon_{10}\rangle+|1\rangle|\varepsilon_{11}\rangle)$$

$$= |+\rangle\frac{|\varepsilon_{00}\rangle+|\varepsilon_{01}\rangle-|\varepsilon_{10}\rangle-|\varepsilon_{11}\rangle}{2} + |-\rangle\frac{|\varepsilon_{00}\rangle-|\varepsilon_{01}\rangle-|\varepsilon_{10}\rangle+|\varepsilon_{11}\rangle}{2}$$

$$= |+\rangle|\varepsilon_{-+}\rangle + |-\rangle|\varepsilon_{--}\rangle$$

In error-check, these parameters should be constrained by the error rates of X-basis and Z-basis,

$$e_z = \langle\varepsilon_{01}|\varepsilon_{01}\rangle = \langle\varepsilon_{10}|\varepsilon_{10}\rangle$$

$$e_x = \langle\varepsilon_{+-}|\varepsilon_{+-}\rangle = \langle\varepsilon_{-+}|\varepsilon_{-+}\rangle$$

$$= \frac{1-\text{Re}(\langle\varepsilon_{00}|\varepsilon_{11}\rangle+\langle\varepsilon_{00}|\varepsilon_{01}\rangle+\langle\varepsilon_{01}|\varepsilon_{10}\rangle+\langle\varepsilon_{10}|\varepsilon_{11}\rangle)}{2}$$

$$= \frac{1-\text{Re}(\langle\varepsilon_{00}|\varepsilon_{11}\rangle-\langle\varepsilon_{00}|\varepsilon_{01}\rangle+\langle\varepsilon_{01}|\varepsilon_{10}\rangle-\langle\varepsilon_{10}|\varepsilon_{11}\rangle)}{2}$$

$$= \frac{1-\text{Re}(\langle\varepsilon_{00}|\varepsilon_{11}\rangle+\langle\varepsilon_{01}|\varepsilon_{10}\rangle)}{2}$$

It is easy to see that $S(\rho^{ABE})$ is monotonically decreasing with $\Delta_1$ and $\Delta_2$. Therefore, it takes its maximum at $\Delta_1 = 0$ and $\Delta_2 = 1-2e_x-2e_z$. Thus,

$$S(\rho^{ABE}) = 1 + h(\xi)$$

where $\xi = \left(1-\sqrt{(1-2p)^2+(1-2e_x-2e_z)^2[1-(1-2p)^2]}\right)/2$ and $h(x) = -x\log_2 x - (1-x)\log_2(1-x)$ is the binary Shannon entropy. This attack satisfies the following equation:

$$\langle\varepsilon_{00}|\varepsilon_{01}\rangle = \langle\varepsilon_{11}|\varepsilon_{10}\rangle = 0$$

$$\langle\varepsilon_{10}|\varepsilon_{01}\rangle = e_z$$

$$\langle\varepsilon_{00}|\varepsilon_{11}\rangle = 1-2e_x-e_z$$

The maximum of mutual information between Alice and Eve is $I(A:E) \leq h(\xi)$.

References
[1] Jozsa, Richard, and Jürgen Schlienz. "Distinguishability of states and von Neumann entropy." Physical Review A 62.1 (2000): 012301.